\documentclass[pra,twocolumn,aps,floatfix,showpacs,tightenlines,superscriptaddress,amsmath,amssymb,showkeys,10pt,nofootinbib]{revtex4-2}
\usepackage[T1]{fontenc}
\usepackage[latin1]{inputenc}
\usepackage{amsmath,graphicx}
\usepackage{amssymb}
\usepackage{epsfig,graphicx}
\usepackage{mathrsfs}
\usepackage[active]{srcltx}
\usepackage{mdframed}
\usepackage{mathtools}
\usepackage{xcolor}
\usepackage{hyperref}
\usepackage{braket}
\usepackage{gensymb}
\usepackage{enumerate}
\usepackage{epsfig,libertine}
\usepackage[libertine]{newtxmath}
\usepackage{layouts}
\usepackage{hyperref}
\hypersetup{colorlinks=true,urlcolor=red,citecolor=blue,filecolor=magenta,bookmarks=true}
\newcommand{\R}{\mathbb{R}}     %the field of real numbers 
\newcommand{\spc}[1]{\mathcal{#1}}    %the font for a Hilbert space 

\def\d{{\rm d}}    %  the differential in an integral

\def\>{\rangle}                                       % to make a ket
\def\<{\langle}                                        % to make a bra
\def\kk{\>\!\>}                                          % to make a double ket
\def\bb{\<\!\<}                                          %to make a double bra
\newcommand{\st}[1]{\mathbf{#1}}          % boldtype font for vectors
\newcommand{\map}[1]{\mathscr{#1}}            %a linear map on operators
\newcommand{\Tr}{\operatorname{Tr}}          %  the trace

\newtheorem{lemma}{Lemma}
\newtheorem{prop}{Proposition}

\def\Proof{{\bf Proof.~}}
\def\qed{$\blacksquare$ \newline}

\begin{document}

\title{Measuring incompatibility and clustering quantum observables with a quantum switch} 

\author{Ning Gao} 
\affiliation{QICI Quantum Information and Computation Initiative, Department of Computer Science, The University of Hong Kong, Pokfulam Road, Hong Kong}
\author{Dantong Li} 
\affiliation{QICI Quantum Information and Computation Initiative, Department of Computer Science, The University of Hong Kong, Pokfulam Road, Hong Kong}
\author{Anchit Mishra} 
\affiliation{QICI Quantum Information and Computation Initiative, Department of Computer Science, The University of Hong Kong, Pokfulam Road, Hong Kong}
\author{Junchen Yan} 
\affiliation{QICI Quantum Information and Computation Initiative, Department of Computer Science, The University of Hong Kong, Pokfulam Road, Hong Kong}
\author{Kyrylo Simonov} 
\affiliation{Fakult\"at f\"ur Mathematik, Universit\"at Wien, Oskar-Morgenstern-Platz 1, 1090 Vienna, Austria}
\author{Giulio Chiribella} 
\affiliation{QICI Quantum Information and Computation Initiative, Department of Computer Science, The University of Hong Kong, Pokfulam Road, Hong Kong}
\affiliation{Quantum Group, Department of Computer Science, University of Oxford, Wolfson Building, Parks Road, Oxford, OX1 3QD, United Kingdom}
\affiliation{Perimeter Institute for Theoretical Physics, 31 Caroline Street North, Waterloo, Ontario, Canada}

\date{\today}

\begin{abstract}
The existence of incompatible observables is a cornerstone of quantum mechanics and a valuable resource in quantum technologies. Here we introduce a measure of incompatibility, called the mutual eigenspace disturbance (MED), which quantifies the amount of disturbance induced by the measurement of a sharp observable on the eigenspaces of another. The MED provides a metric on the space of von Neumann measurements, and can be efficiently estimated by letting the measurement processes act in an indefinite order, using a setup known as the quantum switch, which also allows one to quantify the noncommutativity of arbitrary quantum processes. Thanks to these features, the MED can be used in quantum machine learning tasks. We demonstrate this application by providing an unsupervised algorithm that clusters unknown von Neumann measurements. Our algorithm is robust to noise can be used to identify groups of observers that share approximately the same measurement context. 
\end{abstract} 

\maketitle 

\emph{Introduction.}--- One of the most striking features of quantum mechanics is the existence of incompatible observables.
 Incompatible observables are at the heart of Bohr's notion of complementarity~\cite{bohr1928quantum} and of Heisenberg's uncertainty principle~\cite{heisenberg1985anschaulichen}, and have non-trivial relations with Bell nonlocality~\cite{wolf2009measurements, quintino2014joint} and other forms of nonclassicality~\cite{uola2014joint, tavakoli2020measurement, debievre2021complete, DeBievre2022, Xu2022}. In addition to their foundational relevance, they play center stage in quantum information technologies~\cite{quintino2014joint, skrzypczyk2019all, invitation},  
 where quantum incompatibility serves as a  resource~\cite{resource_theory, DAriano2022, Buscemi2022}, in a similar way as quantum entanglement and coherence.

Several measures of incompatibility have been proposed in the past years,   including robustness to noise (defined as the minimum amount of  noise that has to be added to a set of incompatible observables in order to make them compatible)~\cite{busch2013comparing, heinosaari2014maximally}, sensitivity to eavesdropping (defined as the minimum amount of  disturbance that an arbitrary entanglement-breaking channel would induce on a quantum system prepared in an unknown eigenstate of the given observables)~\cite{operational_measure}, and disturbance on the measurement statistics (defined as the  maximum distance between the probability distribution of observable $A$ on a given input state and the probability distribution of  $A$ after a measurement of observable $B$ has been performed on the same input state, with the maximum evaluated  over all possible input states)~\cite{mandayam2014measures}. Since all these measures are defined in terms of optimization problems, they  are often hard to compute analytically, and numerical evaluation becomes unfeasible for systems of high dimension.   In addition, 
there is generally no  direct way to estimate these measures from experimental data: in most cases, the best known way to infer the incompatibility of two unknown observables is to perform a full tomography, which is unfeasible  for quantum  systems consisting of many particles. 

In this paper we introduce a measure of incompatibility for sharp observables \cite{busch1996quantum},
 called the Mutual Eigenspace Disturbance (MED). 
The MED quantifies the noncommutativity of the spectral resolutions associated to the two observables, and can be naturally extended to a larger class of noncommutativity measures for  unsharp measurements and general quantum processes. It has a simple closed-form analytical expression and,
 unlike other incompatibility measures, it constitutes a metric on the space of von Neumann measurements, a property that makes it suitable for machine learning applications. The MED and its generalizations to measures of noncommutativity  can be directly estimated using the quantum switch~\cite{chiribella2009beyond, switch}, an operation  that  combines  quantum  processes in a coherently-controlled order.   Estimation of the MED via the quantum switch can be realized with existing technology \cite{procopio2015experimental, rubino2017experimental, goswami2018indefinite, wei2019experimental, guo2020experimental, goswami2020increasing, rubino2021experimental,goswami2020experiments}  and its sample complexity   
  is independent of the dimension of the system, meaning that the number of experiments needed to estimate the MED remains small even for multiparticle systems. 
  
The experimental accessibility of the MED and its metric properties  make it suitable for applications in quantum machine learning. To illustrate the idea, we provide a quantum algorithm that clusters noisy von Neumann measurements based on their mutual compatibility. This algorithm can be used to identify clusters of observers who share approximately the same measurement context \cite{specker1960,kochen1975problem,budroni2021quantum,abramsky2011sheaf}, and thereby could share the same notion of an emergent classical reality \cite{despagnat,schlosshauer2005decoherence,grangier2002contextual,zurek2009quantum,horodecki2015quantum}.    Notably, the observers could be localized in distant laboratories, and the algorithm does not require access to their measurement outcomes, but only to the average evolution associated to their measurement devices.

\emph{MED.}---For  sharp observables, compatibility is equivalent to commutativity \cite{lahti2003coexistence}.  
 Let $A$ and $B$ be  two sharp observables  on a $d$-dimensional quantum system, and let $\st P = (P_i)_{i=1}^{k_A}$ and $\st Q = (Q_j)_{j=1}^{k_B}$ be the projectors on the eigenspaces of $A$ and $B$, respectively. 
  We now introduce a measure of noncommutativity    between $\st P $ and $\st Q $. 
  Imagine that the system is initially in an eigenstate of $A$, say $|\alpha_{i}\>$, picked uniformly at random from the $i$-th eigenspace, with $i$ distributed according to the probability distribution $p_i: = d_{A, i}/d$, where $d_{A,i}$ is the eigenspace's dimension. Then, the system undergoes the canonical (L\"uders) measurement process associated to the observable $B$: with probability $p_B (j)= \<\alpha_i| Q_j |\alpha_i\> $ the measurement yields outcome $j$, leaving the system in the post-measurement state $Q_j |\alpha_i\>/\| Q_j |\alpha_i\>\|$. On average over all outcomes, the density matrix of the system is $\sum_j p_B (j) ~Q_j |\alpha_i\>\<\alpha_i| Q_j/ \| Q_j |\alpha_i\>\|^2 = \map B (|\alpha_i\>\<\alpha_i|)$, where $\map B$ is the dephasing channel defined by the relation $\map B (\rho) := \sum_{j} \, Q_j \rho Q_j$ for arbitrary density matrices $\rho$.  Finally, a measurement of the observable $A$ is performed.   The probability to find the outcome $i$, associated to the original subspace, is $\Tr [ P_i \, \map B ( |\alpha_i\>\<\alpha_i|) ]$. On average, the probability that the system is still found in the original eigenspace is 
\begin{align}
 {\sf Prob} ( A, \map B) := \sum_i\,  p_i \, \int \pi_i ( \d \alpha_i) \, \Tr [ P_i \, \map B ( |\alpha_i\>\<\alpha_i|) ] \, ,
\end{align}
where $\pi_i (\d \, \alpha_i)$ is the uniform probability distribution over the pure states in the $i$-th subspace. 
Explicit calculation yields the expression 
\begin{align}
 {\sf Prob} ( A, \map B) := \frac 1d \,  \sum_{i j}\,   \Tr [ P_i Q_j P_i Q_j]  \, ,
\end{align}
which is related to an extension of the Kirkwood-Dirac quasiprobability distribution \cite{halpern2018quasiprobability,arvidsson2021conditions}.

Note that the role of the projectors $\st P$ and $\st Q$ is completely symmetric. Operationally, this symmetry implies to the relation
\begin{align}\label{symmetry}
  {\sf Prob} ( A, \map B) = {\sf Prob} ( B, \map A) \, ,
\end{align}
where ${\sf Prob} ( B, \map A)$ is the average probability that a randomly chosen state $|\beta_j\>$ from the $j$-th eigenspace of $B$, drawn with probability $q_j: = d_{B,j}/d$ (where $d_{B,j}$ is the eigenspace's dimension), is still found in the same eigenspace after the action of the dephasing channel $\map A (\rho)  := \sum_i P_i \rho P_i$.

Note also that the probabilities in Eq.~(\ref{symmetry}) depend only on the dephasing channels $\map A$ and $\map B$.  Accordingly, we will denote by $D (\map A,\map B) : = 1- {\sf Prob} ( A, \map B) \equiv 1- {\sf Prob} ( B, \map A) $ the average probability of eigenstate disturbance.  We then define the MED of the two observables $A$ and $B$ as \begin{align}\label{eq:MED}
{\rm MED} (\map A, \map B)  : = \sqrt{ D (\map A, \map B)} = \sqrt{  1 - \frac 1 d \sum_{i,j}\Tr [ P_i Q_j P_i Q_j ] } \, .  
\end{align}

The MED exhibits several appealing properties for a measure of incompatibility: 
\begin{enumerate}
\item it is {\em symmetric and nonnegative}, namely ${\rm MED} (\map A, \map B) = {\rm MED} (\map B, \map A) \ge 0$ for every pair of dephasing channels $\map A$ and $\map B$, 
  \item it is {\em faithful}, namely ${\rm MED} (\map A, \map B) > 0$ if and only if $A$ and $B$ are incompatible,
  \item it is {\em decreasing under coarse-graining},  
  \item it is a {\em a metric on von Neumann measurements}, corresponding to observables with non-degenerate spectrum. 
 \item it is {\em robust to noise}: it remains faithful even if one of the channels $\map A$ and $\map B$ is replaced by the evolution resulting from a noisy measurement of the corresponding observable, 
    \item it is {\em maximal for maximally complementary observables}~\cite{kraus_complementary_observables}, that is, observables such that their eigenstates form mutually unbiased bases~\cite{ivonovic1981geometrical,wootters1989optimal}. In general, one has the bound ${\rm MED} (\map A, \map B)  \le \sqrt{1- 1/\min \{ k_A, k_B\}}$, and the maximum value ${\rm MED} (\map A, \map B) = \sqrt{1-1/d}$ and is attained if and only if $A$ and $B$ are maximally complementary. 
 \end{enumerate}
 The proof of the above properties is provided in Appendix~\ref{AppA}. There, we also extend the MED to a broader class of incompatibility measures, given by the expression 
 \begin{align}\label{gmed}
   {\rm MED}_\rho  (\map A, \map B)  : = \sqrt{  1 - {\sf Re} \sum_{i,j}\Tr [ \rho P_i Q_j P_i Q_j ] } \, , 
 \end{align}
 where $\rho$ is a density matrix. The original MED, defined above, corresponds to the case where $\rho$ is the maximally mixed state $I/d$.  
Notably, the generalized MED (\ref{gmed}) is also a metric on von Neumann measurements whenever the density matrix $\rho$ is non-singular, including {\em e.g.} the case where $\rho$ is a thermal state.

\emph{Experimental setup.}---We now provide an experimental setup that can be used to estimate the MED of two observables and, more generally, the amount of noncommutativity between two arbitrary quantum processes.  

The setup  is based on the quantum switch \cite{chiribella2009beyond, switch}, an operation that combines two unknown processes in a coherent superposition of two alternative orders. Previously, the quantum switch was  shown to be able to distinguish between pairs of quantum channels with commuting or anti-commuting Kraus operators~\cite{chiribella2012perfect, Swingle2016}, a task that can be practically achieved with photonic systems \cite{andersson2005comparison, procopio2015experimental}. We now show that the quantum switch can be used to quantify  the amount of noncommutativity of quantum meausurements and, more generally, of arbitrary quantum processes.

Suppose that an experimenter is given access to two black boxes, acting on a $d$-dimensional quantum system. The two black boxes implement two quantum processes $\map C$ and $\map D$ with Kraus representations $\map C(\rho)   =  \sum_i  C_i \rho  C_i^\dag$ and $\map D  (\rho)  = \sum_j  D_j \rho  D_j^\dag $, respectively. The goal of the experimenter is to estimate the noncommutativity of the Kraus operators $(C_i)_i$ and $(D_j)_j$. To this purpose, one can combine  the two boxes in the quantum switch \cite{chiribella2009beyond,switch}, generating a new quantum process $\map S_{\map C,\map D}$ with Kraus operators
\begin{equation}
S_{ij} = C_i D_j \otimes |0\rangle\langle 0| + D_j C_i \otimes |1\rangle\langle 1|
\end{equation}
where $\{|0\>,  |1\>\}$ are basis states of a control qubit. 
The action of the channel $\map S_{ \map C,\map D}$ on a generic product state $\rho \otimes \omega$ is 
\begin{align}\label{switch}
    \nonumber \map S_{\map{C},\map{D}}  (\rho \otimes \omega ) =  & \frac{1}{4} \sum_{ij} \Bigl( \{C_i, D_j\} \rho \{C_i, D_j\}^\dagger \otimes \omega  \\
    \nonumber &+  \{C_i, D_j\} \rho [C_i, D_j]^\dagger \otimes \omega Z \\
    \nonumber &+ [C_i, D_j] \rho \{C_i, D_j\}^\dagger \otimes Z\omega \\
    &+ [C_i, D_j] \rho[C_i, D_j]^\dagger \otimes Z\omega Z \Bigr),
\end{align}
where  $[C_i,D_j]:=  C_iD_j  -  D_j  C_i$ denotes the commutator,  $\{C_i,D_j\}:  =   C_i  D_j  +  D_j C_i$ denotes the anti-commutator, and $Z:  = |0\>\<0| -|1\>\<1|$.

To estimate the noncommutativity of $\map C$ and $\map D$, the experimenter can initialize the control qubit in the maximally coherent state $\omega= |+\>\<+|$, apply the quantum channel $\map S (\map C,\map D)$,  and  measure the control system in the Fourier basis $\{|+\>, |-\>\}$, with $|\pm \> := (|0\> \pm |1\>)/\sqrt 2$.  Using Eq. (\ref{switch}), one can see that the probability of the outcome ``$-$'' is 
\begin{align}\label{p-}
 p_- =  \frac{1}{4} \sum_{ij} \Tr \left(  \rho~   \Big| [C_i, D_j] \Big|^2  \right)  \, ,
 \end{align}
where $|O|^2 : = O^\dag O$ denotes the modulus square of an arbitrary operator $O$.

We define the {\em non-commutativity} of two generic quantum processes $\map C$ and $\map D$ relative to the state $\rho$ as 
\begin{align}\label{ncom}
{\rm NCOM}_\rho  (\map C,  \map D)   :   =   \sqrt{2 p_-}    =  \sqrt{ \frac{\sum_{ij} \Tr \left(  \rho~   \Big| [C_i, D_j] \Big|^2  \right) }2 }
\end{align}
(in the special case $\map C= \map D$ and $\rho= I/d$, a related definition was used in \cite{loizeau2020channel} to quantify the degree of non-commutativity of the Kraus operators of a given channel).  It is evident from the definition that ${\rm NCOM}_\rho  (\map C,  \map D)$ is symmetric and non-negative. When the matrix $\rho$ is invertible,  ${\rm NCOM}_\rho  (\map C,  \map D)$ is a faithful measure of non-commutativity, {\em i.e.} ${\rm NCOM}_\rho  (\map C,  \map D)  = 0$ if and only if every Kraus operator $\map C$ commutes with every Kraus operator of $\map D$.    For composite systems, it is possible to show that the noncommutativity between a maximally entangled measurement and product measurement is at least $\sqrt{1-1/d_{\min}}$, $d_{\min}$ being the  dimension of the  smallest subsystem (see Appendix~\ref{AppB}). 

In the special case where  $\map C$ and $\map D$ are two dephasing channels $\map A$ and $\map B$, explicit calculation yields the relation
\begin{align}\label{ncom=med}
{\rm NCOM}_\rho  (\map A, \map B)   =  {\rm MED}_\rho  (\map A, \map B)    \,.
\end{align}
Hence, the MED of two unknown observables can be directly estimated from experimental data. 
   Crucially, the sample complexity of the estimation procedure is independent of the dimension of the system: for a fixed error threshold $\epsilon$ and for every state $\rho$, the estimate of ${\rm MED}_\rho (\map A,\map B)$ can be guaranteed to have error at most $\epsilon$ with probability at least $1 - \delta$ by repeating the experiment for $n= \log{\frac2{\delta} }/(2\epsilon^2)$ times (see Appendix~\ref{AppC}).  
  
 The experimental estimation of the MED is feasible with photonic systems, in particular in the case where the state $\rho$ is maximally mixed, and its preparation can be achieved by generation two-photon Bell  states.  
  The preparation of the maximally mixed state   is  also standard in the DQC1 model of quantum computing \cite{knill1998power} and can be well approximated in other models of quantum computing with highly mixed states \cite{ambainis2000computing}.      
  In Appendix~\ref{AppD} we also discuss the advantages of the quantum switch set up with respect to other ways to estimating the MED.

 Besides providing direct way to the experimental estimation of the MED, the relation with the noncommutativity  also provides an alternative route to its analytical/numerical evaluation. In Appendix~\ref{AppE} we show that the noncommutativity can be equivalently rewritten as 
\begin{align}\label{alternativencomm}
{\rm NCOM}_\rho (\map C, \map D)  =  \sqrt{  1 -   {\sf Re}  \Tr  [ D  \, \check C   \, (  I  \otimes \rho^T)]} \, ,  
\end{align}
where $\check C$ and $D$ are two operators associated to the maps $\map C$ and $\map D$, respectively.   When the operators $\check C$, $D$ and $\rho$ have a suitable tensor network structure, Eq. (\ref{alternativencomm}) provides a way to  efficiently evaluate the noncommutativity, avoiding the sums in Eqs. (\ref{gmed}) and (\ref{ncom}) which may contain an exponentially large number of terms when the system has exponentially large dimension. In Appendix~\ref{AppF} we also show another equivalent expression that reduces the  noncommutativity to  the overlap between two pure states, a task that can be carried out efficiently in a variety of physically relevant cases, including {\em e.g.} matrix product~\cite{Fannes1992, Verstraete2006, PerezGarcia2007} and MERA states~\cite{Vidal2008}.

\begin{figure}[t]
    \centering
    \includegraphics[width=0.5\textwidth]{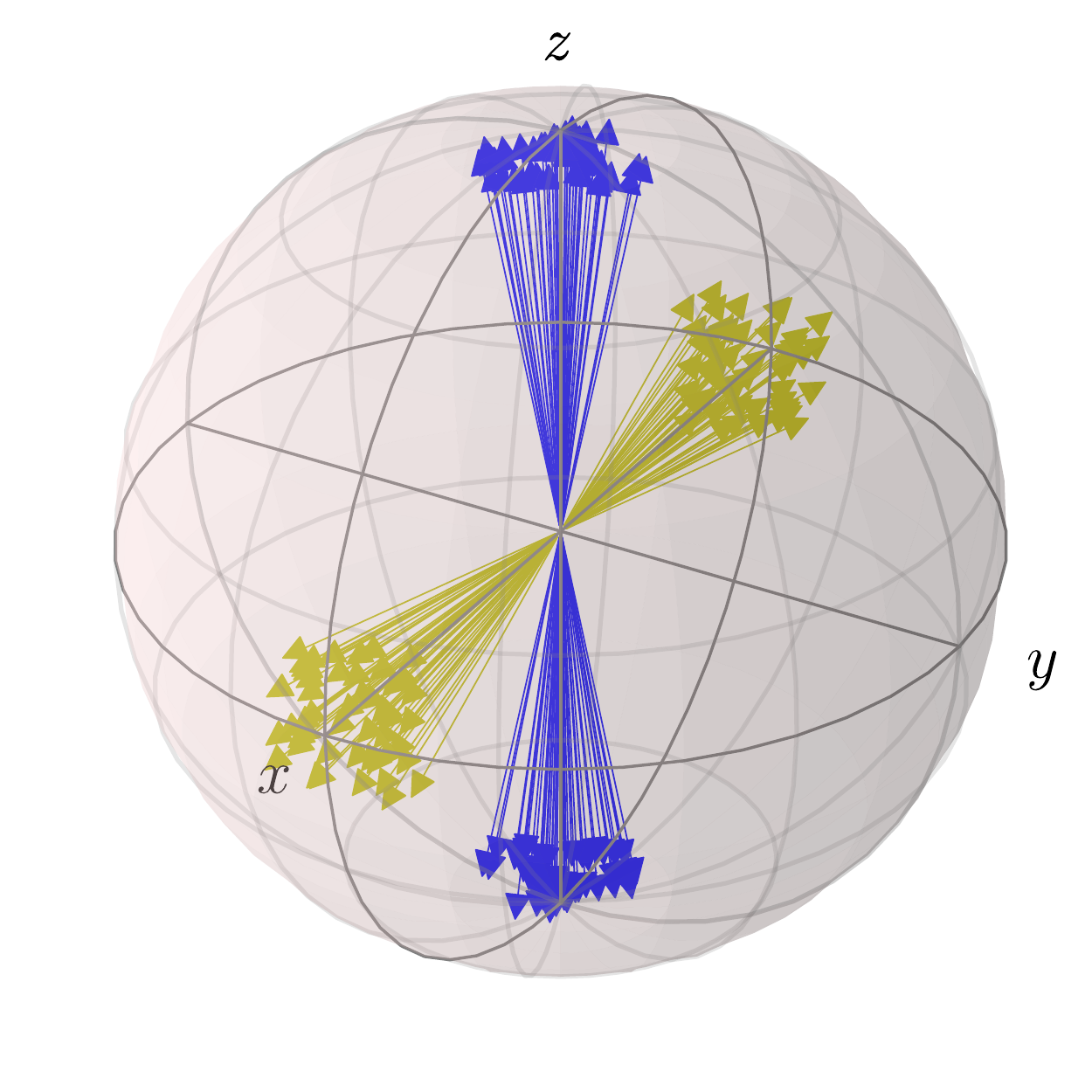}
    \caption{Clustering of $100$ qubit observables, using  a $k$-medoids algorithm based on the values of the MED with $k$-means++ style initial seeding.  The algorithm correctly identifies two clusters, one centered around the Pauli $X$ observable (Bloch vectors in yellow), and one centered around the Pauli $Z$ observable (Bloch vectors in blue). }
    \label{fig:perturbation}
\end{figure}

\begin{figure*}[t]
        \centering
        \minipage{0.32\textwidth}
  \includegraphics[width=\linewidth]{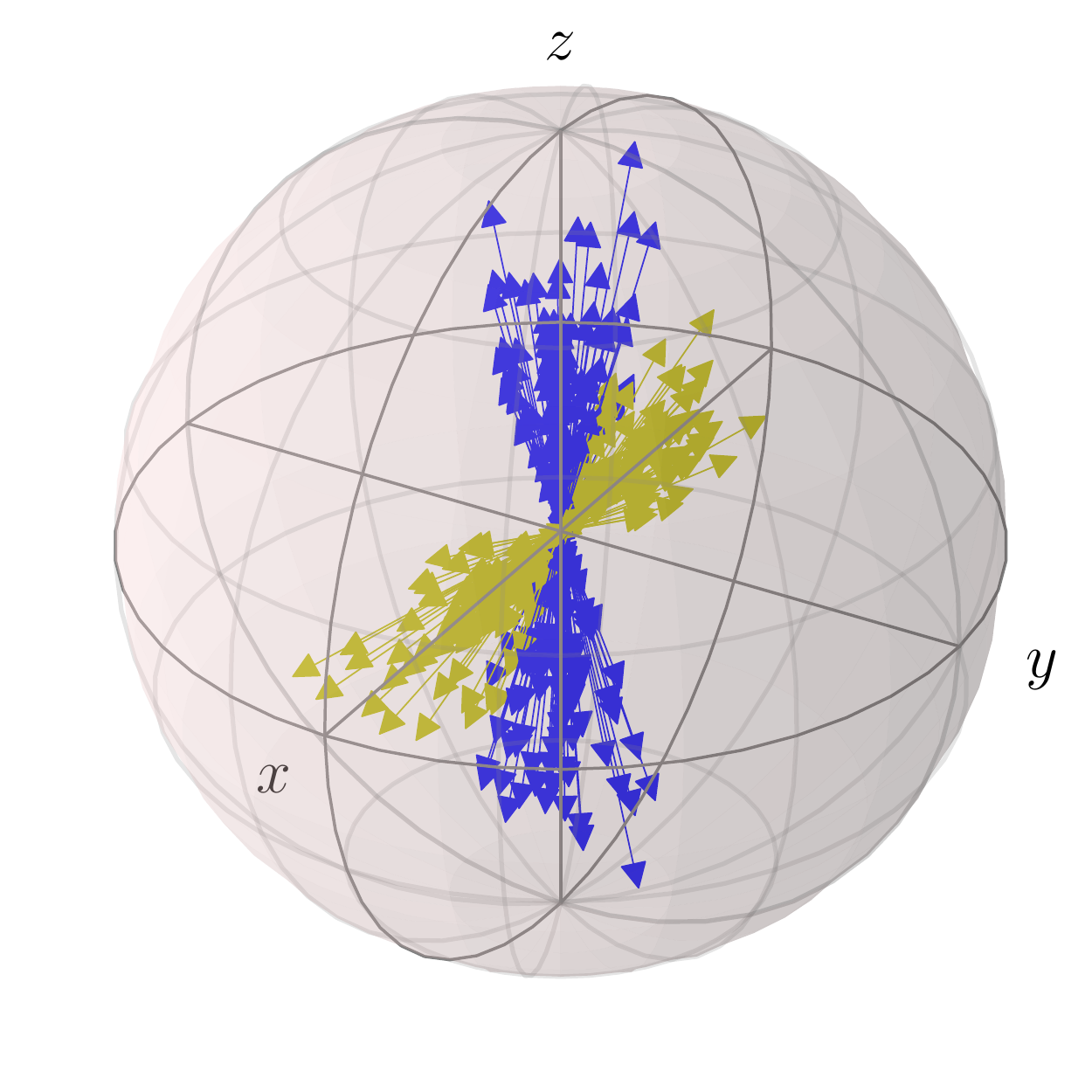}
  \label{fig:awesome_image1}
\endminipage\hfill
\minipage{0.32\textwidth}
  \includegraphics[width=\linewidth]{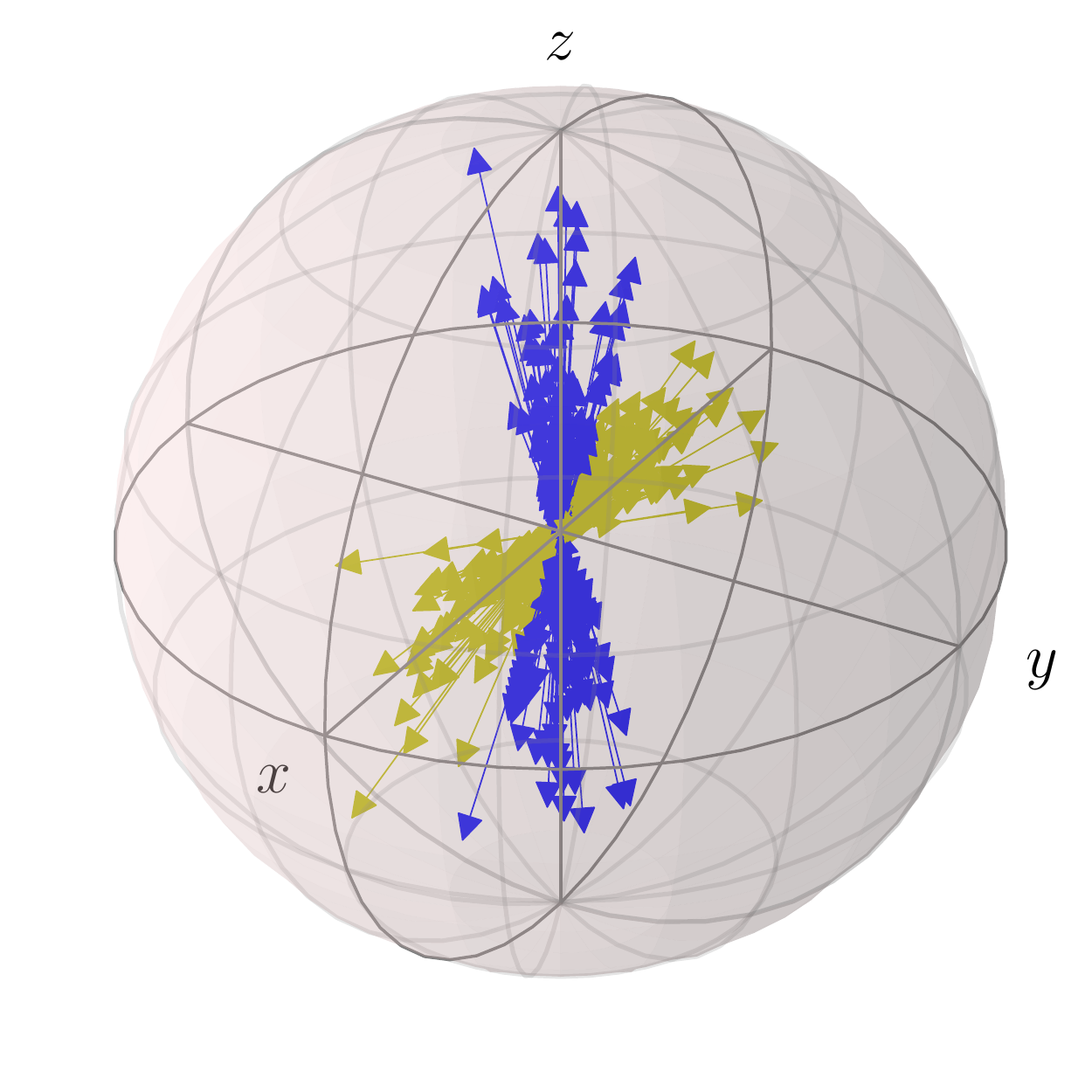}
  \label{fig:awesome_image2}
\endminipage\hfill
\minipage{0.32\textwidth}%
  \includegraphics[width=\linewidth]{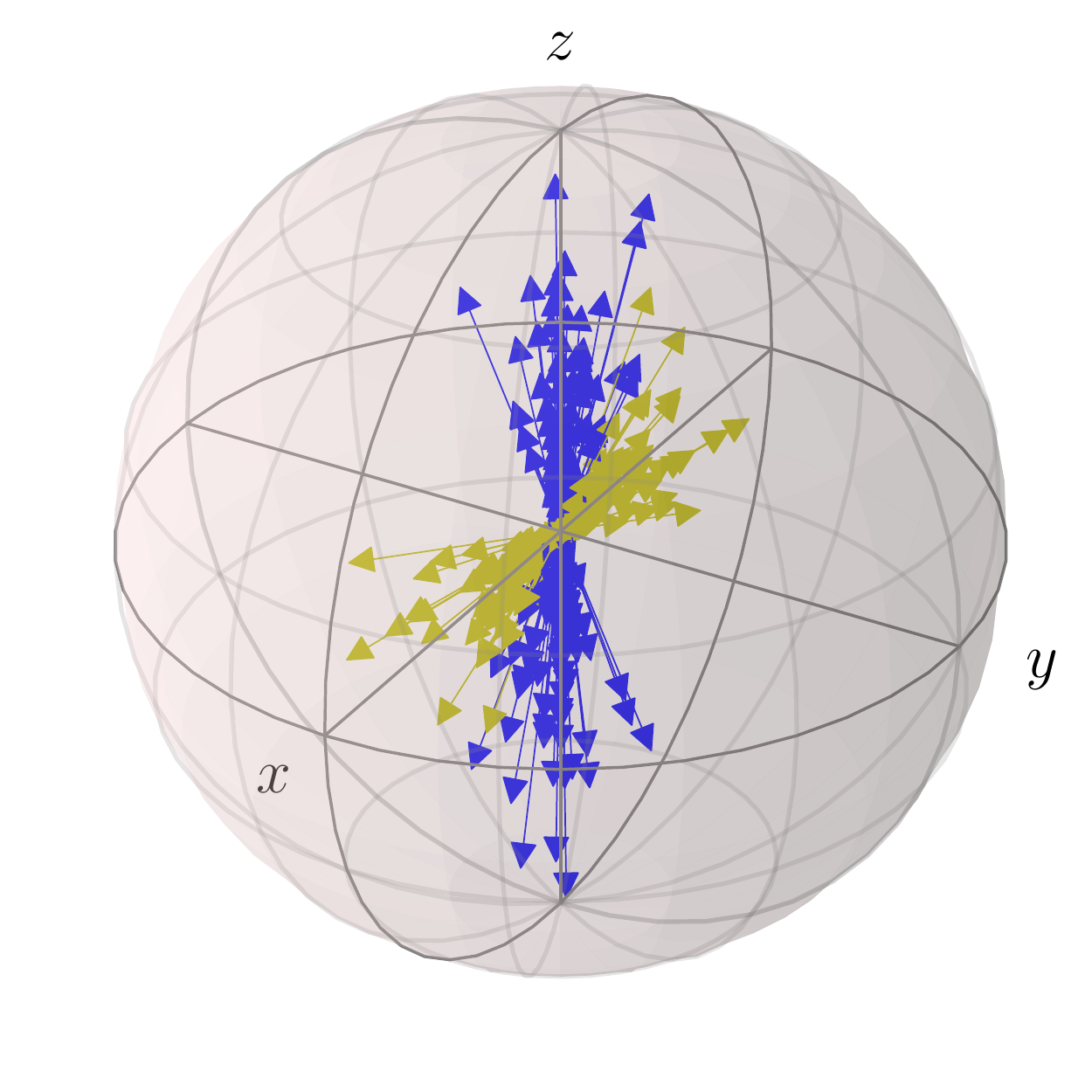}
  \label{fig:awesome_image3}
\endminipage
    \caption{Clustering of $100$ noisy qubit measurements, using  a $k$-medoids algorithm based on the values of the noncommutativity with $k$-means++ style initial seeding.  The algorithm correctly identifies two clusters, centered around the Pauli $X$ and $Z$ observables, respectively. The three plots refer to numerical experiments with maximum noise level $0.25$, $0.5$, and $0.75$, respectively.}
        \label{fig:anisotropic}
\end{figure*}

\emph{Clustering algorithm for quantum observables.}---We now provide a  machine learning algorithm for identifying clusters of observables that are approximately compatible with one another. The algorithm is unsupervised: the learner does not need to be trained with  labelled examples of observables belonging to different clusters.
 
The input of the algorithm is the access to $m$ black boxes, implementing $m$ unknown dephasing channels $\map A_1,\dots, \map A_m$ associated to non-degenerate quantum observables $A_1,\dots,  A_m$.  The quantum part of the algorithm is the estimation of the MED  for every pair of observables. Then, the estimated values of the MED are fed  into a classical clustering algorithm. Here we choose  $k$-medoids clustering with $k$-means++ style initial seeding \cite{scikit-learn, kmeans}.  Compared to  the popular $k$-means method, $k$-medoids  works better  (in terms of convergence)  with arbitrary dissimilarity measures.

To illustrate the algorithm,  we generate $m=100$ random qubit observables, of the form $A_l  =  b^{(l)}_x \,  X  +  b^{(l)}_y \,  Y +   b^{(l)}_z \,  Z $,  $l\in  \{1,\dots,  m\}$,  where $X,Y,Z$ are the three Pauli matrices and $\st b_l=  (b^{l}_x,b^{l}_y,b^{l}_z) \in  \R^3$ is a unit vector (the Bloch vector of the $l$-th observable). The  vectors are generated in the following way: for 50 observables, we start from the Bloch vector $(1,0,0)$  and  apply a rotation by a random angle  $\theta$  with $|\theta| \le 22.5\degree$  about a random rotation axis. For the remaining 50 observables, we start from  the Bloch vector $(0,0,1)$ and apply the same procedure. In this way, the 100 observables are naturally divided into two clusters, as in Figure \ref{fig:perturbation}. 

We  performed a numerical experiment on the classical part of the algorithm, feeding the values of the MED into the $k$-medoids algorithm.  For improved reliability, we repeated the experiment for 50 times, finding that in each repetition all the  $100$ observables are correctly clustered.
Note that, while we fed the algorithm with the exact values of the MED, the robustness of the $k$-medoids \cite{scikit-learn, kmeans} algorithm  implies that the results are robust to errors in the estimation of the MED from actual experimental data.

\emph{Clustering with noisy observables.}---Our clustering algorithm can also be extended  to noisy measurements.  Following \cite{invitation}, the noise is modelled by randomizing the measurement of each ideal observable with a trivial measurement, which produces the same outcome statistics for every possible input state.  Mathematically, this means that the projective measurement $\st P^{(l)}$, associated to the $l$-th observable, is replaced by a non-projective measurement $\st N^{(l) }  = (1 -\lambda_l)  \,  \st P^{(l)}   +   \lambda_l  \,  \st T^{(l)}$, where $\lambda_l$ is the noise probability, and $\st T^{(l)}$ is a trivial measurement, with  POVM operators $T^{(l)}_i  =  p_i^{(l)}\,   I$, for a fixed probability distribution $\st p^{(l)}  =(p^{(l)}_i)$.   For the measurement process associated to the noisy measurement $\st N^{(l)}$, we take quantum instruments with  Kraus operators of the form $N^{(l)}_{i j}   =  \sqrt{  a^{(l)}_{ij}  \,  P_i^{(l)}  +  b^{(l)}_{ij} \,   I }$, where $a^{(l)}_{ij}$ and $b^{(l)}_{ij}$ are arbitrary nonnegative coefficients subject to the constraints $\sum_j   a_{ij}^{(l)}  = 1-\lambda_l$ and   $\sum_j  b_{ij}^{(l)}  = \lambda_l  \,  p_i^{(l)}$ for every $i$ and $l$.  The $k$-medoid algorithm can then be applied, using the noncommutativity (\ref{ncom}) of the noisy channels  $\map N_1, \dots,  \map N_m$ defined by $\map N_l (\rho):  = \sum_{ij}  N_{ij}^{(l)} \rho N^{(l)\dag}_{ij}$ for $l\in  \{1,\dots,  m\}$.      

To test the algorithm, we performed a numerical experiment on 100 randomly generated noisy qubit observables.  
For simplicity, we  chose  isotropic noise~\cite{DallArno2010} and set the noise randomly following a uniform distribution, picking each probability $ p_i^{(l)}$ uniformly at random in the interval $(0,1)$,  subject to the constraint $\sum_i p_i^{(l)} = 1$, $\forall l$.
For the original observables, we generated the Bloch vectors as in the noiseless case.  For the noise, we  first defined a maximum noise level $\eta$ and then we picked a random noise probability  $\lambda_l= \eta  \,  R_l$ where each $R_l$ is chosen  independently, uniformly at random  in the interval $[0,1]$.  The coefficients $a_{ij}^{(l)}$ and $b_{ij}^{(l)}$ are then chosen uniformly at random, subject to the constraints. The experiment has been performed for  $\eta = 0.25$, $0.5$, and $0.75$, and for each setting the $k$-medoids algorithm has been run  50 times.
 The results of the experiment,  plotted on Fig.~\ref{fig:anisotropic},  show that  perfect clustering is still achieved in the presence of noise.

\emph{Conclusions and outlook.}---In this paper we introduced the MED, an experimentally accessible  measure of incompatibility for sharp observables. The MED quantifies the noncommutativity of the projectors associated to a pair of observables,  and can be directly measured without access to the measurement outcomes,  by letting the two measurement processes act in an indefinite order.  Thanks to its properties, the MED can be used in quantum machine learning tasks, such as clustering unknown observables based on their degree of compatibility. 

An interesting direction of future research is to extend the results of this paper to infinite dimensional systems, unsharp observables, and  general quantum channels. For sharp observables with discrete spectrum,  our approach can be easily extended by taking a limit of finite dimensional subspaces. For observables with continuous spectrum, however, the situation is more complex, due to the fact that no repeatable measurement exists~\cite{ozawa1984quantum}.  Regarding the extension to unsharp observables and general channels, one approach is to focus on the noncommutativity,  which can be measured with the quantum switch.  On the other hand, commutativity is not a necessary condition, and an open question is to determine whether there exists a measure of  incompatibility that can be measured experimentally like the MED. 

 Another direction is the extension of the MED from pairs to arbitrary numbers of observables. 
One option would be to  generalize our definition, examining the amount of disturbance on the eigenspace of one observable induced by measurements of the other observables. An appealing feature is that the resulting quantity could be estimated by placing the measurements in a superposition of orders, in a similar way as it was done in our paper for the case of two observables. 

\begin{acknowledgements}
We thank  N. Yunger Halpern,  S. De Bi\`evre, and D. R. M. Arvidsson-Shukur  for  comments on an earlier version of this manuscript, and to L. A. Rozema and M. Antesberger for a discussion on the experimental realization of the proposals in this work.  This research is supported by
the Hong Kong Research Grant Council through Grants No.
17300918 and No. 17307520, and through the Senior Research
Fellowship Scheme SRFS2021-7S02. 
This publication was made possible through the support of the ID$\#$ 62312 grant
from the John Templeton Foundation, as part of the 'The Quantum Information Structure of Spacetime'
  Project (QISS). The opinions expressed in this project are those of the
authors and do not necessarily reflect the views of the John Templeton Foundation. Research at the Perimeter Institute is supported by the
Government of Canada through the Department of Innovation, Science and Economic Development Canada and by the
Province of Ontario through the Ministry of Research, Innovation and Science.
\end{acknowledgements}
 
\appendix

\section{Properties of the   MED and its generalization}\label{AppA}

 Here we  establish the main properties of the MED, including symmetry, nonnegativity, faithfullness, robustness to noise, being a metric for von Neumann measurements, and reaching its maximum value for mutually unbiased bases. Some of the properties will be established for a generalized version of the MED, which  can also be experimentally accessed using the quantum switch.

In the following, we will often consider the {\em generalized MED}  
\begin{align}\label{gmed1} 
{\rm MED}_\rho   (\map A,\map B)   :  =  \sqrt{1- \sum_{i,j}  {\sf Re} \Tr  [  \rho  P_i  Q_j  P_i Q_j]} \,.
\end{align} 
Note that, by definition, one has 
\begin{align}\label{eq:app:MED}
{\rm MED}_{\frac Id}   (\map A,\map B) =  \sqrt{  1 - \frac 1 d \sum_{i,j}\Tr [ P_i Q_j P_i Q_j ] }  \equiv
{\rm MED}  (\map A, \map B)   \, ,
\end{align}
for every pair of observables $A$ and $B$.   In the following we establish a number of properties of ${\rm MED}_\rho$ and ${\rm MED}$.   As it turns out,  some of the properties of  ${\rm MED}_\rho$ require the density matrix $\rho$ to have some properties, such as being invertible, or commuting with the projectors onto the eigenspaces of $A$ and $B$. All these properties are automatically satisfied by the choice $\rho  =  I/d$.

\medskip  

{\bf Symmetry.} By definition, one has   
\begin{align}
\nonumber   {\rm MED}_\rho  (\map A, \map B)  
&  =   \sqrt{1-\sum_{i,j}{\sf Re}   \Tr  [  \rho  P_i  Q_j  P_i Q_j]} \\
\nonumber &  =  \sqrt{1- \sum_{i,j}{\sf Re}   \overline{\Tr  [  \rho  P_i  Q_j  P_i Q_j]}} \\
\nonumber &  =  \sqrt{1- \sum_{i,j} {\sf Re} \Tr  \left[  (\rho  P_i  Q_j  P_i Q_j)^\dag\right]}  \\   
\nonumber &  =  \sqrt{1- \sum_{i,j} {\sf Re} \Tr  [ Q_j  P_i Q_jP_i\rho] } \\
\nonumber &  = \sqrt{1-  \sum_{i,j} {\sf Re} \Tr  [\rho Q_j  P_i Q_jP_i]  } \\
&  =  {\rm MED}_\rho  (\map B,\map A)  \, . \label{eq:Symmetry}
\end{align}
Here the fourth equality follows from the  property  $(XY)^\dag =Y^\dag X^\dag$,  valid for arbitrary operators $X$ and $Y$, along with the fact that the operators $\rho, P_i,$ and $Q_j$ are self-adjoint.  The   fifth equality follows from the cyclic property of the trace.

\medskip 

{\bf Nonnegativity.} The generalized MED can be equivalently expressed as 
\begin{align}\label{eq:rhoswitch}
{\rm MED}_\rho  (\map A, \map B)    =   \sqrt{\frac {  \sum_{i,j} \Tr \Big\{  [  P_i,  Q_j   ]  \rho  \,    [  P_i,  Q_j   ]^\dag \Big\}  }{2}}   
\end{align}
The right-hand side is non-negative, because each of the operators $ [  P_i,  Q_j   ]  \rho  \,    [  P_i,  Q_j   ]^\dag$ is non-negative, and therefore has a non-negative trace. 

\medskip 

{\bf Faithfulness.}  We now show that the generalized MED is faithful for every invertible state $\rho$:  for two arbitrary dephasing channels  $\map A$ and $\map B$, ${\rm MED}_\rho  (\map A, \map B)  >  0$ if and only if the observables $A$ and $B$ are incompatible. 

The proof is based on  Eq. (\ref{eq:rhoswitch}), which shows that ${\rm MED}_\rho  (\map A, \map B)$ is non-negative and equals to zero if and only if $\Tr \big\{  [  P_i,  Q_j   ]  \rho  \,    [  P_i,  Q_j   ]^\dag \big\} =0$ for every $i$ and every $j$.  By the cyclicity of the trace, this condition holds if and only if 
$\Tr \big\{   \sqrt \rho  \,    [  P_i,  Q_j   ]^\dag  [  P_i,  Q_j   ]  \sqrt \rho  \big\} =0$ for every $i$ and every $j$.  Moreover, since each operator $ \sqrt \rho  \,    [  P_i,  Q_j   ]^\dag  [  P_i,  Q_j   ]  \sqrt \rho$ is positive, its trace is zero if and only if the operator itself is zero.  Finally, since $\rho$ is invertible, the condition $ \sqrt \rho  \,    [  P_i,  Q_j   ]^\dag  [  P_i,  Q_j   ]  \sqrt \rho =  0$  holds if and only if   $[  P_i,  Q_j   ]^\dag  [  P_i,  Q_j   ] =0$, or equivalently, if and only if $[  P_i,  Q_j   ] =0$.  Summarizing, the condition ${\rm MED}_\rho  (\map A, \map B)  =  0$ is equivalent to the condition  $[  P_i,  Q_j   ] =0 \, ,\forall i,j$, which is equivalent to the compatibility of the observables $A$ and $B$.

\medskip     

{\bf Monotonicity  under coarse-graining.}  Here we show that  ${\rm MED}_\rho  (\map A, \map B)$ is non-increasing under coarse-graining, provided that the density matrix $\rho$ commutes with the measurement that is being coarse-grained:  
 \begin{prop}
  If the state $\rho$ commutes with the observable $A$, then one has  ${\rm MED}_\rho  (\map A,\map B)  \ge {\rm MED}_\rho  (\map A',\map B)$ whenever $\map A'$ is the dephasing channel associated to a coarse-grained measurement of $A$. 
 \end{prop}

\Proof  Let $\st P'  :=  (P'_k)_{k=1}^{o'}$ be a coarse-graining of the measurement $\st P =  (P_i)_{i=1}^o$, meaning that there is a surjective function $f:  \{1,\dots,  o\} \to \{1,\dots,  o'\}$ such that 
 \begin{align}\label{pkprime} 
 P_k'  =  \sum_{i:  f(i)  =  k}  \,    P_i \,.
 \end{align}  
 Note that, by definition, one has the operator inequality 
 \begin{align}\label{coarse}
 P_i  \le  P_{f(i)}' \,.
 \end{align} 
Now, suppose that the density matrix $\rho$ commutes with the projectors $(P_i)_{i=1}^o$.  In this case, one has the relation
\begin{align}\label{positivewithrho}
\nonumber \Tr [  \rho P_i  Q_j P_{i'} Q_j ] &  = \Tr [  ( \sqrt \rho P_i \sqrt \rho)  \,    ( Q_j P_{i'} Q_j) ]  \\
&  \ge 0 \,,  
\end{align}
valid for every pair of indices $i$ and $i'$.  Here, the last inequality follows from the fact that the operators $\sqrt \rho P_i \sqrt \rho$ and $Q_j P_{i'} Q_j$  are positive semidefinite, and therefore the trace of their product is  non-negative.

In particular, Eq. (\ref{positivewithrho}) implies that $\Tr [  \rho P_i  Q_j P_{i} Q_j ] $ is a real number.  Hence, the real part in the definition of the generalized MED can be dropped, and one has 
\begin{align}\label{medrhoAB}
{\rm MED}_\rho  (\map A, \map B)   =  1-  \sum_{i,j}    \Tr [  \rho P_i  Q_j P_i Q_j ]  \, .
\end{align}
Since $\rho$ commmutes also with the projectors $(P'_k)_{k=1}^{o'}$, we also have the relation 
\begin{align}\label{medrhoA'B}
{\rm MED}_\rho  (\map A', \map B)   =  1-  \sum_{k,j}    \Tr [  \rho P_k'  Q_j P_k' Q_j ]  \, ,
\end{align}
where $\map A'$ is the dephasing channel associated to the coarse-grained measurement $(P_k')_{k=1}^{o'}$

At this point, it is easy to see that ${\rm MED}_\rho  (\map A', \map B) \le {\rm MED}_\rho  (\map A, \map B)$.    Indeed,  one has the bound
\begin{eqnarray}
\nonumber \sum_{k,j}   \Tr [  \rho    \,  P_k'  \,  Q_j  P_k'  Q_j] &=& \sum_{k,j} \sum_{\substack{i: f(i) = k \\ i': f(i') = k}}  \Tr [\rho  \,  P_i  \,  Q_j  P_{i'}  Q_j] \\
\nonumber &=& \sum_{i,j} \Tr [ \rho \, P_i  \,  Q_j  P_i  Q_j] \\ 
\nonumber &&  + \sum_{k,j}    \sum_{\substack{i: f(i) = k \\ i': f(i') = k \\ i \neq i'}}  \Tr [ \rho  \, P_i  \,  Q_j  P_{i'}  Q_j] \\
&\geq& \sum_{i,j} \Tr [\rho  \,  P_i  \,  Q_j  P_i  Q_j], \label{CG_Proof}
 \end{eqnarray}
where the last inequality  follows from Eq. (\ref{positivewithrho}). The inequality ${\rm MED}_\rho  (\map A', \map B) \le {\rm MED}_\rho  (\map A, \map B)$ then follows by combining Eq. (\ref{CG_Proof}) with Eqs. (\ref{medrhoAB}) and (\ref{medrhoA'B}).  \qed

\medskip 

{\bf  Metric on von Neumann measurements.} We now show that the generalized MED based on an invertible state is a metric on the set of rank-one projective measurements. 

\begin{lemma}\label{lem:norm}
For every pair of dephasing channels $\map A$ and $\map B$ associated to rank-one measurements, one has the expression 
\begin{align}
    \label{eq:metricMED}
    {\rm MED}_\rho (\map A, \map B)  =  \frac 1{\sqrt{2}}  \,   \|  ( {\tau_{\map A}-\tau_{\map B}})   \,  \,  (\sqrt \rho \otimes I)  \|_2 \, ,   
\end{align}
where  $\|  X\|_2  =  \sqrt{ \Tr[X^\dag X]}$ is the Hilbert-Schmidt norm, and $\tau_{\map A}$ and $ \tau_{\map B}$ are the Choi operators of $\map A$ and $\map B$, given by
\begin{align}\label{ChoiA}
 \tau_{\map A} := \sum_{m,n}    \map A  (|m\>\<n|) \otimes |m\>\<n|   =  \sum_{i}   P_i \otimes \overline P_i, \\
 \tau_{\map B} :=  \sum_{m,n}    \map B  (|m\>\<n|) \otimes |m\>\<n|    =  \sum_{j}  Q_j\otimes \overline Q_j\, ,
\end{align}
with $ \overline X$ denoting the complex conjugate of a matrix $X$. 
\end{lemma}

\Proof{Let us define the product 
\begin{align}
\<X,  Y\>_\rho  :=  \Tr[ (\rho \otimes I)  \, X^\dag  Y  ] \,.
\end{align}
For every self-adjoint operator $\tau$, we have the relation 
\begin{align}
\nonumber \|  \tau  \,  (\sqrt \rho  \otimes I) \, \|_2^2  &  = \Tr [(\sqrt \rho \otimes I )   \tau^2  (\sqrt \rho \otimes I ) ]  \\
\nonumber  &  = \Tr [(\rho \otimes I )   \tau^2  ]  \\
&  =  \<  \tau, \tau\>_\rho  \, .
\end{align}
Using the two equations above, we obtain 
\begin{align}
    \label{eq:metricExpansion}
    ||  ({\tau_{\map A}-\tau_{\map B}})\,  (\sqrt \rho\otimes I)  \,||_2^2 =& \<(\tau_{\map A}-\tau_{\map B})  , (\tau_{\map A}-\tau_{\map B})   \>_\rho   \nonumber\\
  \nonumber   =&\<\tau_{\map A}   , \tau_{\map A}   \>_\rho + \<\tau_{\map B}    , \tau_{\map B}   \>_\rho \\
    & \quad  - \<\tau_{\map A}, \tau_{\map B}\>_\rho   -   \<\tau_{\map B}, \tau_{\map A}\>_\rho   \, .
\end{align}
Note that one has
\begin{align}
 \nonumber   \<\tau_{\map A}, \tau_{\map A}\>_\rho  
 &  = \Tr [ (\rho \otimes I)  \,  \tau_{\map A}^2 ]\\
\nonumber &  = \Tr [ (\rho \otimes I)  \,  \tau_{\map A} ]\\
 \nonumber &  = \sum_i  \Tr [\rho  \,  P_i]  \, \times \, \Tr [\overline P_i] \\
  \nonumber &  = \sum_i  \Tr [\rho  \,  P_i]  \,  \\
    &   = 1
 \, ,
\end{align}
where the second equality follows from the fact that $\tau_{\map A}$ is a projector, the third equality follows from  Eq. (\ref{ChoiA}), the fourth equality follows from the fact that the measurement is rank-one (and therefore $\Tr [\overline P_i]=1$ for every $i$),  and the fifth equality follows from the normalization condition $\sum_i P_i  =  I$.
Similarly, we have 
\begin{align}
 \nonumber   \<\tau_{\map B}, \tau_{\map B}\>_\rho  
 &  = \Tr [ (\rho \otimes I)  \,  \tau_{\map B}^2 ]\\
 \nonumber  &  = \Tr [ (\rho \otimes I)  \,  \tau_{\map B} ]\\
 \nonumber &  = \sum_j \Tr [\rho  \,  Q_j]  \,  \times \,  \Tr [  \overline Q_j]  \\
  \nonumber &  = \sum_j \Tr [\rho  \,  Q_j]   \\
    &   = 1
 \, .
\end{align}
For the remaining terms, we have 
\begin{align}
    \label{eq:metricAB}
    \<\tau_{\map A}, \tau_{\map B}\>_\rho &= \Tr[(\rho\otimes I)  \,  {\tau_{\map A}^{\dagger}\tau_{\map B}}] \nonumber\\
    &  =\sum_{i,j}\Tr [\rho  \,   P_i  Q_j   ]  \, \times \,    \Tr [  \overline P_i  \, \overline Q_j]
    \nonumber\\
       &  =\sum_{i,j}\Tr [\rho  \,   P_i  Q_j   ]  \, \times \,    \Tr [   P_i  \,  Q_j]
    \nonumber\\
    &= \sum_{i, j}   \Tr [\rho  \,P_i Q_j P_i Q_j ] \,,
\end{align}
where the last equality follows from the fact that the operators $P_i$ and $Q_j$ are rank-one. 

Similarly, we have
 \begin{align}
    \label{eq:metricBA}
    \<\tau_{\map B}, \tau_{\map A}\>_\rho &= \Tr[(\rho\otimes I)  \,  {\tau_{\map B}^{\dagger}\tau_{\map A}}] \nonumber\\
    &  =\sum_{i,j}\Tr [\rho  \, Q_j  P_i   ]  \, \times \,    \Tr [\overline Q_j \overline P_i]
    \nonumber\\
   &  =\sum_{i,j}\Tr [\rho  \, Q_j  P_i   ]  \, \times \,    \Tr [ Q_j P_i]
    \nonumber\\
  \nonumber   &=\sum_{i,j}  \,    \Tr  [  \rho \, Q_j  P_i Q_j P_i]\\
    &    =\sum_{i,j}  \,    \overline{\Tr  [  \rho \,  P_i Q_j P_i Q_j]}.
\end{align}

Hence, 
\begin{align}
    \label{eq:metric2}
    \|{\tau_{\map A}-\tau_{\map B}}\| ^ 2 
    =& 2 - 2  {\sf Re}  \,\sum_{i, j} {\Tr[\rho P_i Q_j P_i Q_j]} 
    \nonumber\\
    =& 2 \,  \left[{\rm MED}_\rho(\map A, \map B)\right]^2.
\end{align}

Eq.~(\ref{eq:metric2}) shows that ${\rm MED}_\rho(\map A, \map B)$ coincides with the Hilbert-Schmidt norm  $||    ({\tau_{\map A}-\tau_{\map B}})  \, (\sqrt \rho \otimes I)   ||_2$ up to a constant factor. 
\qed

\medskip  

\begin{lemma}
 For every invertible density matrix $\rho$, ${\rm MED}_\rho $ is a metric on the space of von Neumann measurements.   
\end{lemma}

\Proof To prove that ${\rm MED}_\rho $ is a metric, we need to show that 
\begin{enumerate}
\item ${\rm MED}_\rho  (\map A,  \map B)  ={\rm MED}_\rho  (\map B,  \map A) $ for every $\map A$ and $\map B$  (symmetry)
\item ${\rm MED}_\rho  (\map A,  \map B)  \ge 0 $ for every $\map A$ and $\map B$, with the equality if and only if $\map A   =  \map B$ (nonnegativity and identity of indiscernibles)
\item ${\rm MED}_\rho  (\map A,  \map C)  \le  {\rm MED}_\rho  (\map A,  \map B)   +  {\rm MED}_\rho  ( \map B, \map C)$ (triangle inequality).  
\end{enumerate}
Symmetry  was established at the beginning of this Supplemental Material, in Eq.~(\ref{eq:Symmetry}).

Nonegativity of the generalized MED was also established earlier in this Supplemental Material in the demonstration of faithfulness of ${\rm MED}_\rho$. When $\rho$ is invertible,    we also  showed that ${\rm MED}  (\map A,\map B)= 0$ implies $[  P{_i,  Q_j}]  =  0 $ for every $i$ and $j$.  
For von Neumann measurements,  $(P_i)_i$ and $(Q_j)_j$ are two maximal sets of rank-one projectors, and  the commutation condition means that  there exists a permutation $\pi  :  \{1,\dots, d\}  \to  \{1,\dots, d\}$ such that $P_i   =   Q_{\pi(i)}$.   In this case,  one has $\map A  (X)  =  \sum_i P_i  X P_i  =  \sum_i Q_{\pi(i)}   X    Q_{\pi  (i)}   =  \sum_j  Q_j  X  Q_j  =  \map B (X)$ for every $d \times d$ matrix $X$.  

Finally, the triangle inequality can be deduced from Eq.~(\ref{eq:metricMED}) of Lemma~\ref{lem:norm} and from the triangle inequality of the Hilbert-Schmidt norm. \qed    }

\medskip  

 {\bf Robustness to noise.}  We have seen that the generalized MED based on an invertible state $\rho$ is a faithful measure of noncommutativity.  We now show that this faithfulness property is preserved even when the ideal projective measurements of $A$ and $B$ are replaced by noisy measurements. Precisely, we consider the noisy scenario where the
  canonical channels $\map A$ and $\map B$ are replaced by quantum channels $\map A'$ and $\map B'$  of the form $\map A'  =  \sum_{i,k} A_{i,k} \rho  A_{i,k}^\dag$ and $\map B'(\rho)  = \sum_{j,l}  B_{j,l} \rho  B_{j,l}^\dag$, with 
  \begin{align}\label{noisy1}
  \sum_k A_{i,k}^\dag A_{i,k} =  (1-\lambda)    \,   P_i  +  \lambda p_i   \,  I 
  \end{align} and 
  \begin{align}\label{noisy2}
  \sum_l B_{j,l}^\dag B_{j,l} =  (1-\mu)  \,   Q_j  +  \mu  \,   q_j\,   I ,
  \end{align}
  with suitable probabilities $\lambda, \mu \in  [0,1]$ and suitable probability distributions $\st p$ and $\st q$.      
In the following, we show robustness under the assumption that at least one of the two channels $\map A'$ and $\map B'$ is self-adjoint (recall that a linear map $\map M$ is self-adjoint if, for every pair of $d\times d$ matrices $X$ and $Y$, one has $\Tr [  X  \map M (Y) ]  =  \Tr [  \map M(X)  Y]$). 

In the noisy case, we consider the noncommutativity 
\begin{align}
{\rm NCOM}_\rho  (\map A',  \map B')     =    \sqrt{ \frac{\sum_{i,j,k,l} \Tr \left(  \rho~   \Big|[A_{i,k}, B_{j,l}] \Big|^2   \right) }2 } \, .
\end{align}
If this quantity is zero, then each of the commutators $[A_{i,k}, B_{j,l}]$ must vanish, i.e. one must have the relations 
\begin{align}\label{comm}
[  A_{i,k},  B_{j,l}]  =  0      \qquad \forall i,j,k,l 
\end{align}
and 
\begin{align}\label{commdagdag}
[  A_{i,k}^\dag,  B_{j,l}^\dag]  =  0      \qquad \forall i,j,k,l  \,,
\end{align}
where the second relation is obtained from the first by taking the adjoint on both sides of the equality sign. 

Note that the above relations  must hold for every possible Kraus decomposition of the channels $\map A'$ and $\map B'$.  If  channel $\map A'$ is self-adjoint, the operators $( A_{i,k})_{i,k}$ also form a Kraus representation, and therefore one must have the relation 
\begin{align}\label{dagcomm}
[  A_{i,k}^\dag,  B_{j,l}]  =  0   \qquad \forall i,j,k,l \, ,
\end{align}
and, taking the adjoint on both sides of the equality 
\begin{align}\label{commdag}
[  A_{i,k},  B_{j,l}^\dag]  =  0   \qquad \forall i,j,k,l \, .
\end{align}
Similarly, if channel $\map B'$ is self-adjoint, the above relations must hold.   Using Eqs.  (\ref{noisy1}), (\ref{noisy2}, (\ref{comm}),  (\ref{commdagdag}), (\ref{dagcomm}), and (\ref{commdag}), we then obtain  
\begin{align}
(1-\lambda)  (1-\mu) \,  [  P_i  ,  Q_j  ]  &  =  \sum_{k,l} [  A_{i,k}^\dag A_{i,k} ,  B_{j,l}^\dag  B_{j,l}] =  0 \qquad \forall i,j \, .  
\end{align}

Hence, if the noncommutativity of  the channels $\map A'$ and $\map B'$ is zero, then the ideal measurements $(P_i)_i$  and $(Q_j)_j$ must be compatible, for every value of the noise parameters $\lambda$ and $\mu$ except in the trivial case $\lambda =1$ or $\mu  = 1$, in which the original measurements are replaced  by white noise.

\medskip 

 {\bf Maximality for maximally complementary observables.} We now show the inequality ${\rm MED} (\map A, \map B)   \le \sqrt{1-  1/\min \{  k_A,  k_B\}}$, where $k_{A}$  ($k_B$) is the number of projectors in the spectral decomposition of the observable $A$  ($B$).  The maximum value is given by ${\rm MED} (\map A, \map B)  = \sqrt{1-1/d}$ and attained if and only if $A$ and $B$ are maximally complementary \cite{kraus_complementary_observables} or in other words, their POVM operators are rank-one projectors onto the  basis vectors of two  mutually unbiased bases.

The proof uses a series of lemmas.  
\begin{lemma}
\label{lem:diagonalization}
Let $A, B$ be $n \times n$ Hermitian matrices: If $A$ and $B$ are positive semi-definite, then $AB$ is diagonalizable and has non-negative eigenvalues. 
\end{lemma}  
\Proof  The proof can be found in Corollary 7.6.2(b) of Ref. \cite{horn_johnson_matrix_analysis}.  \qed 

\begin{lemma}\label{lem:boundmed}
For every pair of observables $A$ and $B$, one has the bound ${\rm MED} (\map A,  \map B) \le \sqrt{1 -  \frac 1  {k_B}}$ where $k_B$ is the number of distinct eigenvalues of $B$.  The equality holds only if $\Tr [  P_i Q_j]$ is independent of $j$.  
\end{lemma}

{\bf Proof.} Using the definitions~(\ref{gmed1}) and~(\ref{eq:app:MED}), we focus on a term $\sum_j{\Tr[P_iQ_jP_iQ_j]}$, fixing thereby the index $i$. If $P_i$ is a projector of rank $r_i$, then
\begin{equation}
    \operatorname{rank}(P_i Q_j) \le  r_i.
\end{equation}
Both $P_i$ and $Q_j$ are orthogonal projectors, they are Hermitian and positive semi-definite operators. Using Lemma~\ref{lem:diagonalization}, we can diagonalize the operator $P_i Q_j$ as
\begin{align*}
    P_iQ_j = X_{ij}\Delta_{ij} X_{ij}^{-1},
\end{align*}
where $\Delta_{ij}$ is a diagonal matrix with non-negative entries.  Let $r_{ij}$ be the rank of $\Delta_{ij}$ and assume that the first $r_{ij}$ diagonal entries of $\Delta_{ij}$, denoted by $\lambda_{ij, 1}, \cdots \lambda_{ij, r_i}$, are non-zero.  Then, we can write
\begin{equation}
        \Tr[P_iQ_j] = \sum_{s = 1}^{r_{ij}}{\lambda_{ij,s}}.
\end{equation}        
Taking into account the relation $ \sum_{j=1}^{k_B} \Tr[P_iQ_j]   =  \Tr[P_i]  =:  r_i$, we find
\begin{equation}\label{eq:ProjConstraint}
        \sum_{j=1}^{k_B} \sum_{s = 1}^{r_{ij}}{\lambda_{ij,s}} =   r_i.
\end{equation}        
On the other hand,
\begin{equation}\label{eq:SquaredEigenvalues}
        \sum_{j=1}^{k_B} \Tr[P_iQ_jP_iQ_j] = \sum_{j=1}^{k_B} \sum_{s = 1}^{r_{ij}}{(\lambda_{ij,s})^2}.
\end{equation}        
The minimum of (\ref{eq:SquaredEigenvalues}) under the constraint  (\ref{eq:ProjConstraint}) can be computed with the method of Lagrange multipliers.  
The coefficients that minimize   (\ref{eq:SquaredEigenvalues})  are given by 
\begin{align}\label{optimallambda}
    \lambda^{\min}_{ij,s}   =  \frac{  r_i}{\sum_{l=1}^{k_B}   r_{il}}  \qquad \forall j \in  \{1,\dots,  k_B\}, \,  \forall s  \in  \{  1\dots,  r_{ij}\}\, .
\end{align}
Plugging the optimal  coefficients into the right hand side of Eq. (\ref{eq:SquaredEigenvalues}) we then obtain   
\begin{equation}
        \sum_{j=1}^{k_B} \Tr[P_iQ_jP_iQ_j] \geq \frac{r_i^2}{ \sum_{l=1}^{k_B}   r_{il}}\, .
\end{equation}       
Now,  recall that $r_{il}  =  \operatorname{rank}  (P_i  Q_l)  \le  \operatorname{rank} (P_i)  =  r_i$. Plugging this relation into the previous inequality, we obtain  
\begin{equation}\label{eq:LBSquaredEigenvalues}
        \sum_{j=1}^{k_B} \Tr[P_iQ_jP_iQ_j] \geq \frac{r_i}{ k_B}\, .
\end{equation}       
Finally, summing over $i$ yields the lower bound 
\begin{equation}\label{eq:LowerBound1}
        \sum_{i=1}^{k_A} \sum_{j=1}^{k_B} \Tr[P_iQ_jP_iQ_j] \geq \frac{d}{k_B} \, .
\end{equation}        
Hence, we obtained 
\begin{align}
 \nonumber  {\rm{MED}}(\map A, \map B)   &=  \sqrt{1 - \frac{\sum_{j=1}^{k_B} \Tr[P_iQ_jP_iQ_j]}d   } \\
  &\le  \sqrt{1-  \frac 1{k_B}} \, .
\end{align}

A necessary condition for achieving the equality is that the eigenv0alues $\lambda_{ij,s}$ depend only on $i$ (and not on $j$ and $s$), cf. Eq. (\ref{optimallambda}).    This condition implies in particular that the sum  $\sum_s\,  \lambda_{ij,s}  =  \Tr [P_i Q_j]$ is independent of $j$.  \qed

\begin{lemma}
\label{lem:boundmed2}
For every two observables $A$ and $B$, one has the bound ${\rm MED} (\map A, \map B)   \le \sqrt{ 1-\frac 1{\min\{k_A,  k_B\}}}$.  The equality is achieved only if 
\begin{align}\Tr [P_i Q_j]=   \frac d{k_A  k_B}   \qquad \forall i  \in  \{1,\dots,  k_A\} \, ,\forall j\in  \{1,\dots,  k_B\}\,.
\end{align}
\end{lemma}
    
\Proof Lemma \ref{lem:boundmed} implies the upper bound  ${\rm MED} (\map A, \map B)  \le \sqrt{ 1-\frac 1{k_B}}$. Moreover, the symmetry of the MED  yields the condition ${\rm MED} (\map A, \map B)  = {\rm MED} (\map B, \map A)  \le \sqrt{ 1-\frac 1{k_A}}$ (the inequality following again from Lemma \ref{lem:boundmed}).  Hence, one has the bound  ${\rm MED} (\map A, \map B)   \le \sqrt{ 1-\frac 1{\min\{k_A,  k_B\}}}$.  By Lemma  \ref{lem:boundmed}, the equality holds only if  $\Tr[P_i Q_j] $ is independent of $j$, and   only  $\Tr [Q_j P_i]  \equiv  \Tr  [P_i Q_j]$ is independent of $i$.  In summary, it is necessary that $\Tr [P_i Q_j]$ is constant, say $\Tr [  P_i  Q_j  ]  =c $ for some constant $c$ and for every $i$ and $j$.  The value of the constant can be obtained from the condition 
\begin{align}
\nonumber d  &  =  \Tr [I]  \\
\nonumber   &  =  \sum_{i= 1}^{k_A}  \sum_{j=1}^{k_B}\,  \Tr [ P_i Q_j]  \\
\nonumber   &  =  \sum_{i= 1}^{k_A}  \sum_{j=1}^{k_B}  \,c   \\
&  = k_A \, k_B \, c \, ,
\end{align}
 which implies $c=  d/(k_A k_B)$. 
 \qed

\begin{lemma}
For every  two observables $A$ and $B$, one has the bound ${\rm MED} (\map A, \map B)   \le \sqrt{ 1-\frac 1d}$ and the equality holds if and only if  $A$ and $B$ are non-degenerate and their eigenvectors form two mutually unbiased bases. 
\end{lemma}

\Proof The inequality ${\rm MED} (\map A, \map B)   \le \sqrt{ 1-\frac 1d}$ is immediate from Lemma \ref{lem:boundmed2} and from the fact that $\min\{k_A, k_B\}$ is at most $d$. Let us now determine when the equality sign is attained. A first necessary condition is that $\min\{k_A, k_B\}  =  d$,  which implies $k_A =  k_B = d$, that is,  both observables $A$ and $B$ are nondegenerate.   The necessary condition in Lemma \ref{lem:boundmed2} then becomes $\Tr [P_i Q_j]   =  1/d\, ,  \forall i,j$.  Hence, $P_i$ and $Q_j$ must be projectors on two vectors from two mutually unbiased bases.

Conversely, suppose that $(|\alpha_i\>)_{i=1}^d$
 and $(|\beta_j\>)_{j=1}^d$ are two mutually unbiased bases, and that $(P_i)_{i=1}^d$ and $(Q_j)_{j=1}^d$ are the corresponding rank-1 projectors.   Then, we have 
 \begin{align}
 \Tr [  P_i Q_j P_i Q_j]    =  \left|  \< \alpha_i  | \beta_j\> \right|^4   =   \frac 1{d^2}  \qquad \forall i,j  \, .   
 \end{align}
Hence, the MED of the corresponding observables is  
\begin{align}
\nonumber {\rm MED}  (\map A, \map B)    
&=  \sqrt{  1  -  \frac 1 d  \, \sum_{i,j}   \Tr [P_i Q_j P_i Q_j] }\\
&=  \sqrt{  1  -   \frac 1d    } \,.
\end{align}
In summary, the MED reaches the maximum value ${\rm MED}  (\map A,\map B)   =  \sqrt{ 1 -  \frac 1 d} $ if and only if $A$ and $B$ are nondegenerate observables associated to mutually unbiased bases. 
 \qed

 \medskip 
\section{Noncommutativity between entangled and product measurements}  \label{AppB}

The mathematical description of a measurement process, including both the measurement statistics and the post-measurement states, is provided by a quantum instrument,  that is,  a collection of completely positive, trace non-increasing maps  $(\map C_i)_{i=1}^k$ such that the sum $\sum_{i=1}^k \map C_i$  is trace-preserving.   Here the index $i$ labels the possible measurement outcomes, which occur on an input state $\rho$ with probability $p(i|\rho)  =  \Tr [\map C_i (\rho)]$, leaving the system in the post-measurement state $\rho_i  =  \map C_i (\rho)/ \Tr[\map  C_i (\rho)]$.  The average evolution due to the measurement is then given by the quantum channel $\map C:  = \sum_{i=1}^k \map C_i$.   

We  define the noncommutativity between two  instruments as the noncommutativity of the corresponding average evolutions:  
\begin{align}
   {\rm NCOM}_\rho  \left( (  \map C_i)_{i=1}^{k_A}  \, ,  \, (  \map D_j)_{j=1}^{k_B}   \right) :  =   {\rm NCOM}_\rho    ( \map C,  \map D )  \, ,
\end{align}
 with $\map C:  = \sum_{i=1}^{k_A} \map C_i$ and $\map D:  = \sum_{j=1}^{k_B} \map D_j$.   

We now provide a lower bound on the noncommutativity between a maximally entangled measurement and a product measurement.  Consider  the case of a bipartite system, consisting of two subsystems $S_1$ and $S_2$, of dimensions $2\le d_1\le d_2$.  By ``maximally entangled measurement" we mean a quantum instrument  $(\map C_i)_{i=1}^{k_A}$ of the form $\map C_i  (\rho)  =C_i \rho C_i^\dag$ with $C_i  =  \lambda_i   |\Psi_i\>\<\Phi_i|$ where $|\Phi_i\>$ and $|\Psi_i\>$ are maximally entangled states, and $0\le\lambda_i\le 1$ is a suitable coefficient.     By a ``product measurement'', we mean an instrument   $    (\map D_j)_{j=1}^{k_B}$ of the form  $\map D_j  (\rho)  = D_j \rho D_j^\dag $ with   $D_j  = \mu_j   \, |\gamma_j\>\< \alpha_j| \otimes |\delta_j\>\<\beta_j|$, where $|\alpha_j\>$ and     $|\gamma_j\>$  ($|\beta\>$ and $|\delta_j\>$) are pure states of system $S_1$  ($S_2$), and $0\le \mu_j\le 1$ is a suitable coefficient.   (Note that we are not assuming that the instrument $(\map D_j)_{j=1}^{k_B}$ can be realized by local operations and classical communication. In other words,  our notion of product measurement corresponds to fine-grained separable instruments \cite{chitambar2014everything}, which are not necessarily realizable through local operations and classical communication).

The noncommutativity of the instruments $(\map C_i)_{i=1}^{k_A}$   and $(\map D_j)_{j=1}^{k_B}$  is given by
\begin{align}
\nonumber {\rm NCOM}_{\rho}   \Big(   (\map C_i)_{i=1}^{k_A},    (\map D_j)_{j=1}^{k_B}  \Big)    &  =  \sqrt{ \frac{\sum_{i,j}  \Tr \Big(  \rho \,  \big|[ C_{i},  D_{j} ]\big|^2 \Big)}2  }\\
  &  =  \sqrt{ 1 -   {\sf Re} \left[ \sum_{i,j}   \Tr[  C_{i}^\dag D_{j}^\dag  C_{i}  D_{j}   \rho   ] \right]   }\, .
\end{align}
Here we consider the case where the state $\rho$ is maximally mixed, namely $\rho  = I_1/d_1\otimes I_2/d_2$. In this case, one has  the bound
\begin{align}
\nonumber    {\sf Re} \left[ \sum_{i,j}   \Tr[  C_{i}^\dag D_{j}^\dag  C_{i}  D_{j}  ] \right]   & \le   \sum_{i,j}  \left| \Tr[  C_{i}^\dag D_{j}^\dag  C_{i}  D_{j}  ] \right|\\
\nonumber &  =  \sum_{i,j}   \lambda_i^2  \,     \big|  \<\Phi_i|  D_j  |\Phi_i\>  \,  \<\Psi_i|  D_j^\dag  |\Psi_i\> \big|\\
&  \le \frac 1{d_1} \sum_{i,j}   \lambda_i^2  \, \mu_j \,      \big|  \<\Phi_i|  D_j  |\Phi_i\>  \,  \big| \, ,
\end{align}
the last bound following form the expression   $D_j  = \mu_j   \, |\gamma_j\>\< \alpha_j| \otimes |\delta_j\>\<\beta_j|$   and from the fact that the state $|\Psi_i\>$ is maximally entangled. 

Using the polar decomposition $D_j  =  U|D_j|$, we then have the bound 
\begin{align}
\nonumber   \sum_i  \lambda_i^2  \,     \big|  \<\Phi_i|  D_j  |\Phi_i\> \big|  & =  \sum_i    \lambda_i^2  \,     \big|  \<\Phi_i|  ~  U|D_j| ~ |\Phi_i\> \big|  \\
 \nonumber & \le \sqrt{   \sum_i   \lambda_i^2  \,\< \Phi_i | ~ |D_j|~  |\Phi_i\>}  \\
   \nonumber & \quad \times \sqrt{  \sum_i  \lambda_i^2\, \< \Phi_i | U |D_j| U^\dag |\Phi_i\>}\\
  \nonumber  &  =  \Tr [|D_j|]\\
    &  =  \mu_j\, ,
\end{align}
the second to last equality following from the completeness relation $ \sum_{i=1}^{k_A} \lambda_i^2  |\Phi_i\>\<\Phi_i |=  I_1\otimes I_2$, implied by the normalization of the instrument $(\map C_i)_{i=1}^{k_A}$.    

Summarizing, we obtained the bound 
\begin{align}
 \nonumber {\sf Re} \left[ \sum_{i,j}   \Tr[  C_{i}^\dag D_{j}^\dag  C_{i}  D_{j}  ] \right] &\le       \frac 1{d_1}  \sum_j \mu_j^2 \\
   &  =  \frac 1{d_1}  ~   (d_1d_2) \, ,\end{align}
where the last equality follows  by taking the trace on both sides of the completeness relation $ \sum_{j=1}^{k_B} \mu_j^2  |\alpha_j\>\<\alpha_j|\otimes |\beta_j\>\<\beta_j|=  I_1\otimes I_2$, implied by the normalization of the instrument $(\map D_j)_{j=1}^{k_B}$.   

Hence, the noncommutativity is lower bounded as 
\begin{align}
\nonumber  & {\rm NCOM}_{\frac{  I_1\otimes I_2}{d_1 d_2}}    \Big(   (\map C_i)_{i=1}^{k_A},    (\map D_j)_{j=1}^{k_B}  \Big)   \\
\nonumber & \quad  =  \sqrt{ 1 -   {\sf Re} \left[ \sum_{i,j}   \frac 1{d_1d_2}  \Tr[  C_{i}^\dag D_{j}^\dag  C_{i}  D_{j}      ] \right]   }\\
  &\quad  \ge \sqrt{ 1 - \frac 1 {d_1}  }   \, .
\end{align}
The above bound can be immediately extended to the multipartite case, by considering a bipartition of the system into the lowest dimensional sybsystem and all the remaining ones.  

 \medskip 
 
\section{Sample complexity of (generalized) MED estimation} \label{AppC}
 Our protocol provides a direct estimate of the MED in terms of the probability  that a Fourier  basis measurement on the control system yields outcome ``$-$''.   The probability can be estimated from the frequency of the outcome ``$-$'' in a number of repetitions of the experiment. We now show that the  number of repetitions is independent of the system's dimension and scales inverse polynomially with the desired level of accuracy.    

The proof is standard, and is provided here just for completeness. The result of the Fourier basis measurement defines a Bernoulli variable with outcomes $k  \in  \{  0,1\}$. Here, outcome 1 corresponds to the outcome corresponding to the basis vector $|-\>$, while outcome 0 corresponds to the basis vector $|0\>$.   The probability mass function  of this Bernoulli variable is $f(k  ; p_-)$ defined by 
\begin{align}
  f(  - ;  p_-)   =  p_-  \qquad  {\rm and} \qquad     f(  + ;  p_-)   = 1- p_-  \,.
\end{align}
If the experiment is repeated for $n$ times, the outcomes are a sequence of independent Bernoulli  variables $X_1, X_2, ..., X_n$, with each variable distributed according to the probability mass function $f (k;  p_-)$.   
The sum of these variables, denoted by $S = \sum_{i} X_i$, is distributed according to the binomial distribution  $B(s;n, p_-)   :  = p_-^{s}   (1-p_-)^{n-s}   \begin{pmatrix} n \\ s
\end{pmatrix}$. Then,  Hoeffding's inequality~\cite{Hoeffding1963} implies that the empirical frequency $S/n$  is close to $p_-$ with high probability, namely
\begin{align*}
	\forall \epsilon > 0, P\left(\left|\frac{S}{n} - p\right| < \epsilon\right) \geq 1 - 2e^{-2\epsilon^{2}n},
\end{align*}
Hence, the minimum number of repetitions of the experiment needed to  guarantee that the empirical frequency has probability at most $\delta$ to deviate from the frequency by at most $\epsilon$ is
\begin{align}
	n(\epsilon, \delta) =   \left \lceil -\frac{1}{2\epsilon^{2}}\log{\frac{\delta}{2}}  \right \rceil \, .
\end{align}
This expression shows that the sample complexity is independent of the dimension of the system under consideration. In particular, the sample complexity does not increase when the MED is measured on multiparticle systems, in contrast with the sample complexity of process tomography which increases exponentially with the number of particles.  

\medskip

\section{Comparison with other experimental schemes} \label{AppD}
 
 A na\"ive way to estimate the MED is  to characterize the channels $\map A$ and $\map B$ via process tomography \cite{chuang1997prescription,poyatos1997complete,leung2003choi,d2001quantum,dur2001nonlocal,altepeter2003ancilla}, or the projective measurements $\st P$ and $\st Q$ via measurement tomography \cite{d2004quantum}, and then use Eq. (\ref{eq:MED}).  However, process and measurement tomography requires a number of experimental settings that grows polynomially with the dimension, and therefore exponentially in the number of particles for multiparticle quantum systems.

To avoid the exponential complexity of tomography, one needs  a direct measurement protocol whose sample complexity is independent of the system's dimension, and, therefore, also on the number of particles. A  possible approach is to use  the operational scheme that motivated  the  definition of the MED.   This scheme, described in the main text, involves the initialization  of the system in a random eigenstate of observable $A$.  Practically, this can be achieved by preparing the maximally mixed state, and then applying  a  measurement of the observable $A$.  The estimation of the MED would then proceed by performing 
  a   measurement of the observable $B$, and finally, another  measurement of the observable $A$.  The total probability that the two measurements of $A$ give equal outcomes is equal to ${\sf Prob} (A, \map B)  = 1 - {\rm MED}  (\map A, \map B)^2 $. In this case, the complexity of estimating the MED does not grow exponentially with the system's size. 
  
The main difference between the above scheme and the scheme using the quantum switch is that  the above scheme   requires   access to the outcomes of two measurements of the observable $A$.  The ability to estimate the incompatibility without access to the outcomes offers an advantage in situations where the experimenter  wants to discover the incompatibility of two observables measured by two parties in their local laboratories. In this case, the quantum switch scheme allows the experimenter to estimate the incompatibility/noncommutativity in a black box fashion, by sending input states to the two laboratories and observing their outputs states, without any access to the outcomes generated inside the laboratories.

\section{Proof of Eq. (11) in the main text}  \label{AppE}

From Eq. (9) in the main text, we have 
\begin{align}
  \nonumber  {\rm NCOM}_\rho   (\map C,  \map D)   &  =  \sqrt{  \frac { \sum_{i,j}  \Tr  \Big(  \rho   \,   \big|[  C_i  , D_j ]\big|^2\Big)}{2}}\\
  \nonumber &  = \sqrt{  1  -  {\sf Re}  \left[ \sum_{i,j}   \Tr \left(  C_i  D_j  \rho  C_i^\dag D_j^\dag\right) \right]} \\
 \nonumber   &  =   \sqrt{  1  -  {\sf Re}  \left[ \sum_{j}   \Tr \left(  \map C  (D_j  \rho) ~ D_j^\dag\right) \right]} \\
   &  =  \sqrt  {1  -  {\sf Re}  \left[  \sum_j  \bb \, D_j \,  |  \, \map C  (D_j\rho) \, \kk \right] }  \, ,
\end{align}
where we used the double-ket notation $|X  \kk :   =   \sum_{m,n}    \<  m|  X  |n\>  \,  |m\>\otimes |n\>$ for an arbitrary operator $X$, and the property $\bb X |  Y\kk  =  \Tr  [  X^\dag Y]  =  \Tr [Y  X^\dag]$, valid for arbitrary $X$ and $Y$ (in our case, $X  = D_j $ and $Y  = \map C  (D_j\rho)$).  

Now, consider the operator $\check C$ defined through   the relation~\cite{chiribella2009realization} 
\begin{align}\label{defcheck}
    \check C    |X\kk  :  =  |\map C (X)  \kk  \, ,  \qquad  \forall X   \, .
\end{align}
 Using this definition,  the noncommutativity can be expressed as 
\begin{align}
  \nonumber  {\rm NCOM}_\rho   (\map C,  \map D)   &  =  \sqrt  {1  -  {\sf Re}  \left[  \sum_j  \bb \, D_j \,  |  \,\check C  \,  |    D_j \rho   \, \kk \right] } \\
\nonumber   &  =  \sqrt  {1  -  {\sf Re}  \left[  \sum_j  \bb \, D_j \,  |  \,\check C  \,      (I \otimes \rho^T)  \,   |D_j   \, \kk \right] } \\
\nonumber   &  =  \sqrt  {1  -  {\sf Re}  \Tr  \left[ \left( \sum_j |D_j   \, \kk \bb \, D_j \,  | \right) \,\check C  \,      (I \otimes \rho^T)  \,    \right] } \\
 &  =  \sqrt  {1  -  {\sf Re}  \Tr  \left[   D  \,\check C  \,       (I \otimes \rho^T)  \,    \right] }\, ,
\end{align}
where $D  :  =  \sum_j |D_j\kk\bb D_j|  =  (\map D\otimes \map I)  (|I\kk\bb I|)$ is the Choi operator of $\map D$, with $\map I$ being identity channel.  

In particular, when the state $\rho$ is maximally mixed, the noncommutativity takes the simple expression  
\begin{align}
    {\rm NCOM}_{\frac Id}   (\map C,  \map D)   
  =  \sqrt  {1  -   \frac{ {\sf Re}  \Tr  \left[    D  \,\check C      \right] }d }\, .
\end{align}

The operator $\check C$ is in one-to-one correspondence with the map $\map C$. Explicitly, Eq. (\ref{defcheck}) implies the explicit expression
\begin{align}
\check C    =  \sum_i  C_i\otimes \overline C_i \, ,
\end{align}
where $\overline C_i$ is the complex conjugate of the matrix $C_i$. 
The operator $\check C$ can be equivalently expressed in terms of its Choi operator via the relation   
  \begin{align}\label{checkfromchoi}
\check C  =  \Big ( I_{\mathrm{out}  , \overline{\mathrm{out}}}  \otimes \bb I_{\mathrm{in}  , \overline{\mathrm{in}}  }|   \Big)  (C_{\mathrm{out , in} }  \otimes I_{\overline{\mathrm{out}}, \overline{\mathrm{in}}})     ~   \Big(   |I_{\mathrm{out}  , \overline {\mathrm{out}}}  \kk   \otimes \mathtt{SWAP}_{\mathrm{in}  , \overline{\mathrm{in}}  } \Big)  \, ,        
\end{align}
where ``$\rm in$'' and ``$\rm out$'' label  the input and output systems of channel $\map C$, respectively,  $\overline{\rm in}$ and $\overline{\rm out}$ are auxiliary systems of the same dimensions of $\rm in$ and $\rm out$, respectively, and ${\tt SWAP}_{{\rm in}  , \overline{\rm in}  }$ is the  swap operator, defined by the relation ${\tt SWAP}_{{\rm in}  , \overline{\rm in}  } | \varphi\>   |\psi\>  =  |\psi\>|\varphi\>$ for every pair of vectors $|\varphi\>$ and $|\psi\>$.   Eq. (\ref{checkfromchoi}) can be verified explicitly using  the expressions $\check C  = \sum_i  C_i\otimes \overline C_i $ and  $C  =  \sum_i  |C_i\kk \bb  C_i|$.

\section{Expression of the noncommutativity of two channels in terms of their unitary realizations } \label{AppF}
Here we provide an alternative expression of the noncommutativity (and therefore of the MED)  in terms of the unitary realizations of the two channels $\map C$ and $\map D$.  Consider two unitary realizations, of the  form 
\begin{align}
\nonumber \map C  (\rho)   &=  \Tr_E   \left[  U_{SE}  \, (\rho_S \otimes |\eta\>\<\eta|_E) \,  U_{SE}^\dag \right]\\
\map D  (\rho)   &=  \Tr_F   \left[  V_{SF}  \, (\rho_S \otimes |\phi\>\<\phi|_F) \,  V_{SF}^\dag \right] \, , 
\end{align}
where $E$ and $F$ are two suitable quantum systems, serving as the environments, $|\eta\>$ and $|\phi\>$ are pure states of $E$ and $F$, respectively, and $U_{SE}$ and $V_{SF}$ are two unitary evolutions between the target system (denoted by $S$) and the environments $E$ and $F$, respectively.  

To compute the action of the channel $\map S  (\map C, \map D)$, we  consider the  quantum switch of the unitary gates $U_{SE} \otimes  I_F$ and $V_{SF} \otimes  I_E$, and then take the partial trace over the environments.  The quantum switch of the unitary gates  $U_{SE} \otimes  I_F$ and $V_{SF} \otimes  I_E$ yields the controlled unitary gate 
\begin{align}
\nonumber  W  &=  (U_{SE} \otimes  I_F  )(V_{SF} \otimes  I_E) \otimes |0\>\<0|_C  \\
  &  \quad  +  (V_{SF} \otimes  I_E)   (U_{SE} \otimes  I_F  )   \otimes |1\>\<1|_C \,,
\end{align}
where the subscript $C$ denotes the control system, and it is  implicitly understood that the Hilbert spaces in the tensor product are suitably arranged according to the subscripts of the corresponding systems. 

Hence, the action of the channel $\map S_{\map C, \map D}$ on the state $\rho \otimes |+\>\<+|$ is given by  
\begin{align}
\nonumber \map S_{\map C, \map D} \, \big(  \rho_S \otimes |+\>\<+|_C \big)   &=  \Tr_{EF}  [W  \, (   \rho_S \otimes |\eta\>\<\eta|_E \\
& \quad \otimes |\phi\>\<\phi|_F \otimes |+\>\<+|_C) \,  W^\dag]\, .
\end{align}

To reduce the above expression to the case where the input state is pure, we take a purification of $\rho$, given by a pure state $|\Psi\>  \in \spc H_S \otimes \spc H_R$, where $R$ is a suitable purifying system.    
Let us define the state 
\begin{align}
\nonumber |\Gamma\>_{SREFC}    &:=  (W \otimes  I_R)   \,  |\Psi\>_{SR}  \otimes   |\eta\>_E  \otimes |\phi\>_{F} \otimes |+\>_C \\
  &= \frac{ | \Lambda_0\>_{SREF}  \otimes |0\>_C  +  | \Lambda_1\>_{SREF}  \otimes |1\>_C   }{\sqrt{2}}\, , 
\end{align}
with 
\begin{align}
  |\Lambda_0\>_{SREF}   :  =&    \Big [  (U_{SE} \otimes  I_F  )(V_{SF} \otimes  I_E)  \otimes I_R  \Big]   \,  |\Psi\>_{SR} \otimes |\eta\>_E \otimes |\phi\>_F \, , \\
  |\Lambda_1\>_{SREF}   :  =&    \Big [  (V_{SF} \otimes  I_E)   (U_{SE} \otimes  I_F  ) \otimes I_R  \Big]   \,  |\Psi\>_{SR} \otimes |\eta\>_E \otimes |\phi\>_F \, .
\end{align}

The probability of the outcome - when the output state is measured on the Fourier basis is then given by  
\begin{align}
\nonumber p_-   &=    \< -  |\,  \Tr_{S}  [\map S_{\map C, \map D} \, \big(  \rho_S \otimes |+\>\<+|_C \big)] \, |-\>  \\
\nonumber   &  =  \<- | \, \Tr_{SREF}  [ |\Gamma\>\<\Gamma] \, | -\> \\
\nonumber   &  =  \left\|    (I_{SREF} \otimes \<- |_C  )  \,  |\Gamma\>_{SREFC}   \,  \right\|^2  \\
\nonumber &  = \frac 14   \left\|     |\Lambda_0\>   -  |\Lambda_1\>     \,  \right\|^2 \\
  &   =  \frac {1   -   {\sf Re} \<  \Lambda_0|\Lambda_1\> }2  \, .
\end{align}

Using the relation, ${\rm NCOM}_\rho  (\map C, \map D)   =  \sqrt{2 p_-}$, we finally obtain 
\begin{align}
    {\rm NCOM}_\rho  (\map C, \map D)   =  \sqrt{1 -   {\sf Re} \<  \Lambda_0|\Lambda_1\> } \, .
    \end{align}
The evaluation of the noncommutativity is reduced to the evaluation of the overlap between the states $|\Lambda_0\>$ and $|\Lambda_1\>$.  When the system $S$ consists of a large number $N$ of particles, its Hilbert space has exponentially large dimension in $N$, and therefore the evaluation of the overlap may  be computationally demanding. However, there exist many relevant situations in which the overlap between two states of exponentially large dimension can nevertheless be computed efficiently, {\em i.e.} in a polynomial number of steps.  This is the case, for example, if the states $|\Lambda_0\>$ and $|\Lambda_1\>$ are matrix product states~\cite{Fannes1992, Verstraete2006, PerezGarcia2007} or MERA states.    Physically, these cases correspond to the situation in which the pure state $|\Psi\>$ and the unitary evolutions $U_{SE}$ and $V_{SF}$ have an appropriate tensor network structure, generated by sequences of local interactions. 

\bibliography{main}

\end{document}